\documentclass[12pt]{article} 
\usepackage{graphicx}
\usepackage{hyperref}
\usepackage{geometry}                		
\usepackage[title]{appendix}

\PassOptionsToPackage{hyphens}{url}\usepackage{hyperref}
\usepackage{amssymb}
\usepackage{pdflscape}
\usepackage{csquotes}

\usepackage{hyperref}
\hypersetup{
    colorlinks=true,
    linkcolor=cyan,
    filecolor=cyan,      
    urlcolor=cyan,
    citecolor=cyan
}

\usepackage{cite}

\usepackage{setspace}

\begin{document}

\title{Online Statistics Teaching and Learning}
\author{Jim Albert, Mine Cetinkaya-Rundel and Jingchen Hu\\ Bowling Green State University, Duke University and Vassar College}

\maketitle

\thanks{Jingchen Hu was supported by The Andrew W. Mellon Foundation Anne McNiff Tatlock '61 Endowment for Strategic Faculty Support at Vassar College and the Liberal Arts Consortium for Online Learning (LACOL).}

\tableofcontents

\section{Introduction}

Statistics is probably one of the most active fields to embrace and engage online teaching and learning. Numerous massive open online courses (MOOC) have been developed and have attracted a great number of online learners. At the time of writing, there are 346 courses and specializations with the keyword ``statistics" on Coursera\footnote{For more information, visit \url{https://www.coursera.org/}}, one of the most popular online learning platform. DataCamp\footnote{For more information, visit \url{datacamp.com}}, a growing online learning platform specializing in data science education, has 159 courses, 22 tracks, and 122 instructors.  A broad survey of online statistics education is described in \cite{mills2011teaching}, and a discussion of building an online statistics curriculum is contained in \cite{young2015challenges}.

For statistics courses at all levels, teaching and learning online poses challenges in different aspects.  Particular online challenges include how to effectively and interactively conduct exploratory data analyses, how to incorporate statistical programming, how to include  individual or team projects, and how to present mathematical derivations efficiently and effectively.

This chapter draws from the authors' experience with seven different online statistics courses to address some of the aforementioned challenges. Section \ref{EDA} is an online exploratory data analysis course taught at Bowling Green State University. Section \ref{BayesShare} is  an upper level Bayesian statistics course taught at Vassar College and shared among 10 liberal arts colleges through a hybrid model. Section \ref{MOOC} is describes a five-course MOOC specialization on Coursera, offered by Duke University.

All of these courses are designed for undergraduate or graduate students with calculus backgrounds. The general aim in this chapter is to provide overviews of these online courses , discuss challenges and approaches, and provide general guidelines for statistics educators interested in online teaching and learning of statistics.

Yang \cite{yang2017} provides an overview of the components of an online statistics course and uses student feedback to gain insight on the particular components that appear effective for learning the statistics material.  Everson and Garfield \cite{EversonGarfield} discuss the use of student discussions in an online statistics course and focus on the types of discussion that appear to facilitate understanding of the statistics concepts.  Dunwill \cite{dunwill2016} provides general comments about the challenges of teaching in an online format and Everson \cite{Everson2009} discusses her experiences in teaching online after teaching face-to-face courses in statistics.

Section 2 gives an overview of the seven online statistics courses focusing on the intended audience and the course content.  Chapters 3 through 6 focus on particular components of the online course, and discuss each course on how it addresses the particular component.  
Chapter 3 discusses the problem of the online course design. How is the instructional content presented and organized, keeping in mind the learning objectives of the course?  Chapter 4 focuses on the use of technology in each course.  A statistics course will typically include the use of some software package.  What types of software are used in each class and how is the software integrated with the learning of the conceptual material?  Chapter 5 describes the different forms of assessment for each course.  What types of assessments such as homework or projects are used and do students have the opportunity to work together on assignments?  Chapter 6 describes how the students interact with the instructor in the course and how students interact with other students.  How does a student get help in the course?  All of the presented online courses don't run smoothly and Chapter 7 describes the challenges that the instructors faced when implementing each class.  Chapter 8 summarizes the general features and challenges of the courses, directed towards the instructor who is developing his or her first online statistics class.

\section{Description of the Online Courses}

\subsection{An Online Course in Exploratory Data Analysis}
\label{EDA}

This online course Exploratory Data Analysis at Bowling Green State University focuses on the principles of exploring data following ideas from John Tukey's EDA book \cite{tukey1977}.  The audience consists of graduate and undergraduate students majoring in statistics and there is a probability prerequisite.  

Generally, the main intent of the course is to describe an exploratory philosophy in the analysis of data.  One does not wish to impose any assumptions such as normal sampling distributions or equality of variances between groups.  Instead, one wishes to explore the data, looking for patterns in distributions and relationships.  There are four ``R"s in EDA that summarize the general philosophy in data exploration.  ``Revelation" means that EDA often uses graphical displays in data discovery.  ``Resistance" means that it is desirable to use statistical methods that are resistant or nonsensitive to outlying values.  ``Reexpression" means that it is sometimes useful to reexpress variables by a nonlinear transformation such as a log or square root.  Last, "Residual" means that one usually wishes to look at the deviations from a statistical fit.

Table \ref{tab:EDA} shows the main units for the EDA course.  The course begins with a discussion of graphical displays and resistant summaries for a single batch of measurement data.  The next general topic is the comparison of batches of measurement data and the use of reexpressions to equalize spreads across batches.  Properties of the Box-Cox power family of transformations \cite{sakia1992} are explored in Unit 4 and this family is used to perform an appropriate reexpression to make a data distribution symmetric.  Scatterplots of two measurement variables are introduced in Unit 5 and Tukey's resistant line is applied as a general method of fitting a straight line to data.  In cases where the scatterplot pattern is nonlinear, a running-median smoother is described as a simple way of smoothing a scatterplot to assess the general pattern. Unit 6 explores a two-way table where one summarizes a measurement variable over two categorical variables.  Median polish is a resistant method of applying an additive fit -- by the use of a logarithmic transformation, this method can also be used to apply a multiplicative fit to this two-way data structures.  The course concludes in Unit 7 by describing methods for binning measurement data, assessing if the histogram has a Gaussian shape, and exploring batches of fraction data.

\begin{table}[htp]
\caption{Unit and lectures for the online EDA course.}
\label{tab:EDA}
\begin{center}
\begin{tabular}{|l|l|} \hline
Unit & Lecture \\ \hline
1. Introduction & 1.1 Introduction to EDA I \\ 
& 1.2 Introduction to EDA II \\ \hline
2. Single Batch & 2.1 Displays \\
& 2.2 Summaries \\ \hline
3. Comparing Batches & 3.1 Boxplots \\
& 3.2 Spread Level Plots \\
& 3.3 Comparing Batches III \\ \hline
4. Transformations & 4.1 Transformations \\
& 4.2 Reexpressing for Symmetry \\
& 4.3 Reexpressing for Symmetry II \\
& 4.4 Transformations Summary \\ \hline
5. Plotting & 5.1 Introduction to Plotting \\
& 5.2 Resistant Line \\
& 5.3 Plotting II \\
& 5.4 Straightening \\
& 5.5 Smoothing \\ \hline
6. Two-Way Analyses & 6.1 Median Polish \\
& 6.2 Plotting Additive Fit \\
& 6.3 Multiplicative Fit \\
& 6.4 Extended Fit \\ \hline
7. Counts and Fractions & 7.1 Binning Data \\
& 7.2 Binning Data II \\
& 7.3 Fraction Data \\ \hline
EDA Project & \\ \hline
\end{tabular}
\end{center}
\label{default}
\end{table}%

\subsection{A Bayesian Statistics Course for Cross-Campus Share}
\label{BayesShare}

The past decades have seen great methodological, computational, and inferential advancement of Bayesian statistics. While Bayesian statistics continues to gain attention and becomes ever more popular among data analysts and researchers, the topic itself is rarely available to students, especially at the undergraduate level. In most liberal arts colleges, with the staffing constraints, offering a topic course on Bayesian statistics can at most be a occasional luxury. More commonly, such a course is not offered at all.

Vassar College had the chance to offer an undergraduate-level Bayesian statistics course in Fall 2016. The extremely positive experience with a small group of motivated students has encouraged the instructor to think beyond the boundary of a physical college location. Vassar is a member of the Liberal Arts Consortium for Online Learning (LACOL)\footnote{LACOL is a partnership of 10 liberal arts colleges in the United States, founded in 2014. By leveraging the power of consortial relationships, LACOL focuses on utilizing and adapting emerging technologies to promote excellent and innovative teaching, learning, and research in the liberal arts. For more information about LACOL, visit \url{http://lacol.net/about-the-consortium/}}. Under the Upper Level Math \& Stats Project\footnote{For more information, visit \url{http://lacol.net/category/collaborations/projects/upper-level-math/}}, starting from Fall 2017, the Bayesian Statistics course at Vassar College is taught locally at Vassar while shared among the LACOL colleges through a hybrid model. 

The Upper Level Math \& Stats Project focuses on sharing upper level mathematics and statistics courses among participating campuses, to supplement existing and probably limited offering while maintain the liberal arts flavor. Vassar's Bayesian Statistics course in Fall 2017 is one of the 3 courses in the pilot study (other two are upper level mathematics courses offered by two other member colleges).

The student audience consists of junior and senior students. The prerequisite includes multivariate calculus, linear algebra, and probability. The textbook is A First Course in Bayesian Statistics Methods by Peter D. Hoff \cite{Hoff2009}, a book mainly used at the graduate level. The instructor intentionally borrows more applied material from Bayesian Cognitive Modeling: A Practical Course, written by Michael D. Lee and Eric-Jan Wagenmakers \cite{LeeWagenmakers2014}, making the course more accessible to undergraduate students. Occasionally, advanced material from Bayesian Data Analysis by Andrew Gelman and others \cite{BDA2013} is used to supplement.

There are three general sections of the course: inference, computation, and applications, with main topics in each section listed in Table \ref{tab:BayesShare}. Through the introduction of one-parameter models such as beta-binomial and normal-normal, the inference section covers the inferential basics. Moving to multi-parameter models such as normal with two unknown parameters, computation techniques are covered. Students are equipped with the skills of writing up Markov chain Monte Carlo (MCMC) sampler when possible, as well as the the use of Just Another Gibbs Sampler (JAGS). Ultimately, through various applications, students are exposed to more advanced models. They are motivated to understand and construct Bayesian models in each application, perform simulation by MCMC with appropriate computation techniques, and answer inferential questions in context.

\begin{table}[htp]
\caption{Section and main topics for the cross-campus shared Bayesian Statistics course.}
\label{tab:BayesShare}
\begin{center}
\begin{tabular}{|l | l|} \hline
Section & Topics \\ \hline
Inference & Bayes theorem, conjugate prior, posterior distribution, \\
& HPD interval, predictive distribution \\ \hline
Computation & Monte Carlo approximation, Markov chain Monte Carlo \\
&  (MCMC), Gibbs sampler, Metropolis-Hastings alogrithm, \\
& MCMC diagnostics, JAGS \\ \hline
Applications & Bayesian hierarchical modeling, Bayesian linear regression, \\
& latent class modeling, Bayesian cognitive modeling \\ \hline
\end{tabular}
\end{center}
\label{default}
\end{table}%

Participating in LACOL's Upper Level Math \& Stats Project is the first time that a Bayesian statistics course is ever shared among colleges from different geographic locations. In addition to motivating and cultivating students' learning of such advanced statistics topics, the instructor needs to redesign an existing face-to-face course to adapt to a hybrid model. Challenges include using software to provide both synchronous and asynchronous access to the lectures, identifying what material is suitable for being moved online, holding online office hours for remote students, coordinating with local faculty liaison from each remote campus, and creating a learning community involving all students, among other things. This course uses R extensively for simulations and data analysis, and how to effectively incorporate R programming through a hybrid instruction model is also challenging.

\subsection{A Five-Course MOOC Specialization: Statistics with R}
\label{MOOC}

Statistics with R is a specialization offered on Coursera (\url{www.coursera.org/specializations/statistics}) comprised of five massive open online courses (MOOCs) designed and sequenced to help learners master foundations of data analysis and statistical inference and modeling. The specialization also has a significant hands on computing component. The target audience is learners with no background in statistics or computing.

The first four courses in the specialization are Introduction to Probability and Data, Inferential Statistics, Linear Regression and Modeling, and Bayesian Statistics. These courses cover exploratory data analysis, study design, light probability, frequentist and Bayesian statistical inference, and modeling. A major focus of all of these courses is hands on data analysis in R; each course features computing labs in R where learners create reproducible data analysis reports as well as fully reproducible data analysis projects demonstrating mastery of the learning goals of each of the courses. The fifth course is a capstone, where learners complete a data analysis project that answers a specific scientific/business question using a large and complex dataset. This course is an opportunity for learners to practice what they learned in the first four courses in the specialization.

Table \ref{tab:modules} shows the modules and associated topics for each of the first four courses in this specialization. Each subsequent course assumes learners have either completed the previous course(s) or have background knowledge equivalent to what is covered in them. Each module is designed to be completed in one week, though learners have the flexibility to extend this if they need to.

\newgeometry{margin=0.5cm} 
\begin{landscape}
\begin{table}
\small{
\begin{tabular}{| p{5.7cm}  p{20cm} |}
\hline
\multicolumn{2}{| l |}{\textbf{Course 1: Introduction to Probability and Data}} \\
\hline
\textit{1.1 Introduction to data} 				& Data basics, observational studies and experiments, sampling and sources of bias, experimental design \\
\textit{1.2 Exploratory data analysis and introduction to inference} & Visualizing data, measures of center and spread, robust statistics, transformations, exploring bi/multivariate relationships, introduction to inference via simulation \\
\textit{1.3 Introduction to probability}			& Independent and disjoint events, conditional probability, Bayes' rule, introduction to Bayesian inference  \\
\textit{1.4 Probability distributions} 			& Normal and binomial distributions, assessing normality \\
\textit{1.5 Data analysis project}				& Exploratory data analysis of data from the Behavioral Risk Factor Surveillance System \\
\hline
\multicolumn{2}{| l |}{\textbf{Course 2: Inferential statistics}} \\
\hline
\textit{2.1 Confidence intervals} 			& Sampling variability and Central Limit Theorem, confidence intervals for a mean, accuracy vs. precision \\
\textit{2.2 Inference and significance} 		& Hypothesis testing for a mean, decision errors, statistical vs. practical significance \\
\textit{2.3 Inference for means}				& t-distribution, inference for a mean and for comparing two or more means, multiple comparisons, bootstrapping \\
\textit{2.4 Inference for proportions} 			& Sampling variability and CLT for proportions, confidence intervals and hypothesis tests for two or more proportions, randomization tests for small samples \\
\textit{2.5 Data analysis project}				& Inference on data from the Behavioral Risk Factor Surveillance System \\
\hline
\multicolumn{2}{| l |}{\textbf{Course 3: Linear regression and modeling}} \\
\hline
\textit{3.1 Linear regression}				& Correlation, residuals, least squares line, prediction and extrapolation \\
\textit{3.2 More on linear regression}			& Outliers. inference for regression, variability partitioning  \\
\textit{3.3 Multiple linear regression}			& Multiple predictors, adjusted $R^2$, collinearity and parsimony, inference for MLR, model selection and diagnostics \\
\textit{3.4 Data analysis project}				& EDA and single and multiple regression for movies data \\
\hline
\multicolumn{2}{| l |}{\textbf{Course 4: Bayesian statistics}} \\
\hline
\textit{4.1 Basics of Bayesian statistics}		& Conditional probabilities and Bayes' rule, diagnostic testing, bayes updating, Bayesian vs. frequentist definitions and inference, effect size and significance \\
\textit{4.2 Bayesian inference}				& From discrete to continuous, elicitation, conjugacy, Gamma-Poisson and Normal-Normal conjugate families, non-conjugate priors, credible intervals, predictive inference \\
\textit{4.3 Decision making}				& Loss functions, minimizing expected loss, Monte-Carlo sampling, prior choice and reference priors, MCMC \\
\textit{4.4 Bayesian regression}				& Bayesian simple and multiple regression, model uncertainty and averaging, decisions under model uncertainty \\
\textit{4.5 Perspectives}					& Interviews with statisticians on how they use Bayesian statistics in their work \\
\textit{4.6 Data analysis project}				& Bayesian inference and regression for movies data \\
\hline
\multicolumn{2}{| l |}{\textbf{Course 5: Statistics with R Capstone}} \\
\hline
\textit{5.1 Exploratory data analysis} 		        &  \\
\textit{5.2 Basic model selection }			&  \\
\textit{5.3 Model Selection and Diagnostics}	&  \\
\textit{5.4 Out of Sample Prediction}			&  \\
\hline
\end{tabular}
}
\caption{Modules and topics for the first four courses in the Statistics with R specialization}
\label{tab:modules}
\end{table}
\end{landscape}
\restoregeometry

Course 1, Introduction to Data, introduces sampling and exploring data, as well as basic probability theory and Bayes' rule. In this course learners examine various sampling methods, and discuss how such methods can impact the scope of inference. In addition, a variety of exploratory data analysis techniques are covered, including using data visualization and summary statistics to explore relationships between two or more variables. Another key learning goal for this course is the use of statistical computing, with R, for hands on data analysis. The concepts and techniques introduced in this course serve as building blocks for the inference and modeling courses in the specialization.

Course 2, Inferential Statistics, introduces commonly used statistical inference methods for numerical and categorical data. Learners learn how to set up and perform hypothesis tests and construct confidence intervals, interpret p-values and confidence bounds, and communicate these results correctly, effectively, and in context without relying on statistical jargon. Building on computing skills they acquired in the previous course, learners conduct these analyses in R.

In Course 3, Linear Regression and Modeling, introduces simple and multiple linear regression. Learners learn the fundamental theory behind linear regression and, through data examples, learn to fit, examine, and utilize regression models to examine relationships between multiple variables. Model fitting and assessment is done in R, and substantial amount of examples are focused on interpretation and diagnostics for model checking.

These first three courses were originally offered as a single, much longer, MOOC, for two years, before being split into shorter courses to be bundled up in a specialization. Course 4, Bayesian Statistics, was added to the sequence at this point, in order to make this introductory specialization more complete by adding a different point of view for approaching statistical analysis. 

Course 4, Bayesian Statistics, introduces learners to the underlying theory and perspective of the Bayesian paradigm and shows end-to-end Bayesian analyses that move from framing the question to building models to eliciting prior probabilities to implementing in R. The course also introduces credible regions, Bayesian comparisons of means and proportions, Bayesian regression and inference using multiple models, and discussion of Bayesian prediction.

The last course in the specialization is a capstone course. The materials provided for this course are designed to serve as a reminder of learning goals of earlier courses or expand on them ever so slightly. A large and complex dataset is provided to the learners and the analysis requires the application of a variety of methods and techniques introduced in the previous courses, including exploratory data analysis through data visualization and numerical summaries, statistical inference, and modeling as well as interpretations of these results in the context of the data and the research question. Learners are encouraged to implement both frequentist and Bayesian techniques and discuss in context of the data how these two approaches are similar and different, and what these differences mean for conclusions that can be drawn from the data.

\section{Online Course Design}

\subsection{EDA Course}

This course was originally taught face-to-face in the classroom where the instructor would introduce and demonstrate the EDA methods in class and the students would work on weekly data analysis assignments.   The  online course was designed to follow the same format as the face-to-face version.

\begin{enumerate}
\item The lecture material for the class is posted online as pdf documents.  Students have had difficulties understanding the material by reading directly from \cite{tukey1977} and so the lecture material seems to be a reasonable substitute for the book material.  The core EDA material is contained in a series of 23 pdf documents.  A particular ``lecture"  motivates and describes the particular EDA method with an illustration using the R programming language. 

\item There are weekly data analysis assignments and no written exams in the course.   It is well-known that exams can be challenging to administer in an on-line format.

\item The work on the assignments is a blend of statistical work such as tables and graphs and interpretation of the results.   These assignments are turned in online by the use of R Markdown documents saved in html format.  
\item All of the R code in the lecture notes is made available to the students by a collection of  R scripts.
\end{enumerate}

\subsection{Bayesian Statistics Course}

\subsubsection{Lectures}

Regular face-to-face lectures are delivered in the classroom with students present at Vassar College. Every lecture is broadcast and recorded by Zoom, a video conferencing software\footnote{For more information about Zoom, visit \url{https://zoom.us/}}. Remote students can join the lecture in real time by a Zoom meeting ID. Otherwise, they can watch the recorded videos after the videos are posted on the same day of the lecture.

\subsubsection{Course material in video form}

This hybrid model has made lectures available in video form. With the flexibility of making videos and the added familiarity of learning through videos on the students' end, some other course material has also been turned into video form. These material has conventionally been available as a word or pdf document.

For example, when an example is not fully developed during lecture due to time constraint, a short video on this example is created and made available to students to review if necessary. As another example, when many students are having problems with the same homework question (based on observation from office hour visits), a short video on providing hint on this homework question is created and made available.

R programming demonstrations are very suited in video form. By watching a video with step-by-step demonstration of programming, students are able to pause when needed, see things in action, and practice along the way. Several R programming videos are created for students in this course.

\subsection{Five-Course MOOC}

\subsubsection{Videos}

Each module includes 7-10 videos roughly 4-7 minutes in length. Most of these videos introduce new concepts and the remaining provide additional examples and worked out problems. The slides that serve as the background in the videos are created in Keynote (Apple's presentation software application) and feature a substantial amount of animations such that text, visualizations, and calculations showed on the slides follow the pace of speech in the videos. Many learners have expressed in their course feedback that these features make the videos more engaging and easier to follow compared to videos in many other MOOCs. Sample videos from the Inferential Statistics course are hosted on YouTube (\url{bit.ly/2LrO6KZ}).

\subsubsection{Learning objectives}

Each module also features a set of learning objectives. A sampling of learning objectives from the Inferential Statistics course are shown below:

\begin{displayquote}
\small{
- Explain how the hypothesis testing framework resembles a court trial. \\
- Recognize that in hypothesis testing we evaluate two competing claims: the null hypothesis, which represents a skeptical perspective or the status quo, and the alternative hypothesis, which represents an alternative under consideration and is often represented by a range of possible parameter values. \\
- Define a p-value as the conditional probability of obtaining a sample statistic at least as extreme as the one observed given that the null hypothesis is true: p-value = P(observed or more extreme sample statistic $|$ $H_0$ true)
}
\end{displayquote}

These learning objectives are constructed using verbs from the revised Bloom's Taxonomy \cite{anderson2001taxonomy} and aim to keep learners organized and focused while watching the videos. The learners are recommended to have the learning objectives handy while watching the videos and revisit sections of the videos and/or suggested readings for any learning objectives that they feel like they have not mastered at the end of the module.

The learning objectives are provided as separate stand-alone documents, and after each batch of related learning objectives are a few simple conceptual questions for learners to check their understanding before moving on.

\subsubsection{Suggested readings and practice}

Suggested readings for the first three courses come from OpenIntro Statistics \cite{diez2014openintro}. This book is free and open source, meaning that learners enrolled in the MOOC do not need to additionally purchase a textbook. The readings are optional as the videos explicitly introduce and cover all required topics for the course, however many learners have reported in their feedback that they really like having a reference book that closely follows the course material. Practice problems are also suggested from the end of chapter exercises in this book.

For the fourth course on Bayesian statistics, readings are suggested from An Introduction to Bayesian Thinking \cite{clyde2018bayesian}. This textbook has been written by the Bayesian Statistics course development team (faculty and PhD students) specifically as a companion to this course and is also freely available on the web.

\section{Use of Technology}

\subsection{EDA Course}

\subsubsection{R package}

A special R package \texttt{LearnEDAfunctions} was written for the class.  This package contains all of the datasets used in the lecture notes and the assignments.  In addition, the package contains special functions for implementing the EDA computations.  For example, the function \texttt{rline} computes Tukey's resistant line and the function \texttt{fit.gaussian} fits a Gaussian comparison curve to histogram data and outputs the rootogram residuals from the Gaussian fit.

\subsubsection{EDA blog, youtube videos}

Different methods have been used to provide weekly communications with the students.  For several iterations of the course, the instructor posted weekly articles on the blog ``Exploratory Data Analysis" (\texttt{https://exploredata.wordpress.com/}).  In a typical post, the instructor would give an example of the weekly EDA concept and give advice on some common problems in applying the interpreting the EDA method.  Blog postings from previous years are made available for the student who wishes to see additional illustrations of the statistical methods.   Since some students expressed preference for learning by watching videos instead of reading notes, the instructor added some videos at the youtube channel \texttt{https://www.youtube.com/user/bayesball2/videos}.  A particular video would show the implementation of a particular EDA method using R and the {\texttt LearnBayesfunctions} package.

\subsubsection{Shiny activities}

In addition to the weekly data analysis assignments, this class also contains several activities where the student uses sliders and other interactive tools in exploring data.  For example, in choosing the ``correct" power of a reexpression, the student can choose a value of the power on a slider and see the immediate impact of that particular reexpression in a graph of the reexpressed data.

\subsection{Bayesian Course}

\subsubsection{Use of technology in lectures}

Instead of writing on the chalkboard or presenting lecture slides on the projector in front of the classroom using a computer, the instructor uses an iPad as a participant of the Zoom meeting, bring up lecture slides inside the Zoom software, share screen to present the slides to the class on the projector, and uses an Apple pencil to write on the slide or a whiteboard. When doing R programming demonstration, the instructor joins the Zoom meeting with a laptop, then share screen of R/RStudio on the laptop to the projector. Zoom records the video from the projector, and all audio from the lecture. Sometimes, the instructor could use a directional mic on the iPad to improve sound capturing.

\subsubsection{Use of videos}

This course extensively uses videos to deliver content.  One of the most creative uses of making videos for this hybrid course is to create guest lectures. Conventionally, guest lectures are delivered in the physical classroom. Now, to include remote students, guest lectures can be created in video form and made available to students online. This practice also saves class meeting time when possible. For example, a guest lecture video by a cognitive science professor at Vassar College is created and used as an introduction to Bayesian hierarchical modeling. Students watch the guest lecture before the class meeting, familiarize with the topic by themselves outside of the class, then when meeting in class, the lecture and discussion can follow from the common ground of the material from the guest lecture directly. This practice also exposes students to applications of Bayesian statistics at various stages of their learning.

\subsection{Five-Course MOOC}

Each module in the course features a computing lab in R. The objective of the labs is to give learners hands on experience with data analysis using modern statistical software, R, as well as provide them with tools that they will need to complete the data analysis projects successfully. 

The statistical content of the labs match the learning objectives of the respective modules they appear in and the application examples (i.e. datasets and research questions) are primarily from social and life sciences. The labs also make heavy use of an R package, \texttt{statsr}, which was designed specifically as a companion for the specialization \cite{statsr}. Two other important aspects of the labs are that (1) they use the \texttt{tidyverse} syntax and (2) they are completed as reproducible R Markdown reports.

The tidyverse is an opinionated collection of R packages designed for data science, meaning that the grammar used in the packages is optimized for working with data -- specifically for wrangling, cleaning, visualizing, and modeling data \cite{wickham2017tidyverse}. The choice for the tidyverse syntax for beginners is rooted in wanting to get learners exploring real and interesting data and build informative and appealing visualizations and draw useful conclusions as much as possible \cite{robinson2017tidyverse}.

R Markdown provides an easy-to-use authoring framework for combining statistical computing and written analysis in one document \cite{xie2018r}. It builds on the idea of \textit{literate programming}, which emphasizes the use of detailed comments embedded in code to explain exactly what the code was doing \cite{Knuth1984}. The primary benefit of R Markdown is that it restores the logical connection between statistical computing and statistical analysis by synchronizing these two parts in a single reproducible report. From an instructional perspective this approach has many advantages: reports produced using R Markdown present the code and the output in one place making it easier for learners to learn R and locate the cause of an error and learners keep their code organized and workspace clean, which is difficult for new learners to achieve if primarily using their R console to run code \cite{cetinkaya2018infrastructure}. Each lab is provided to the learners in an R Markdown template that they can use as a starting point for their lab report. Earlier labs in the specialization include lots of scaffolding, and almost have a fill-in-the-blanks feel to them. As the course progresses the scaffolding in the templates are removed, and by the end of the first course learners are able to produce a fully reproducible data analysis project that is much more extensive than any of their labs. All labs in the specialization are hosted in a publicly available GitHub repository at \href{github.com/StatsWithR/labs}{\url{https://github.com/StatsWithR/labs}}.

\section{Assessments and Engagement}

\subsection{EDA Course}

\subsubsection{Weekly data analysis assignments}

In a typical data analysis assignment, the student works on a particular EDA method using a specific dataset or another suitable dataset chosen by the student.  One challenge for the student is to find a suitable data structure to implement the EDA method.  For example, if the EDA task is to symmetrize a dataset by the use of a power expression, the student needs to find a strongly skewed dataset that could benefit with a reexpression.

\subsubsection{Final project}

In a final capstone project, the students selects his or her own dataset, states some questions of interest, and explores the dataset using several of the EDA methods discussed in class.  The focus of this project is not on the implementation of the methods but rather in the interpretation of the results in light of the questions that were originally posed.

\subsection{Bayesian Course}

\subsubsection{Homework}

Homework is on a biweekly basis. The assignment usually consists of a set of derivation exercises to enhance the understanding of Bayesian methodology, and a set of application-based exercises, which requires the uses of R programming. There are two midterm exams and no final exam.

The teaching assistant for this course holds regular office hours at Vassar. While these office hours are held online too, remote students rarely make use of them. Instead, the instructor meets with the remote students together during a separately scheduled online office hour, also through Zoom.

Homework submission from the remote students is done through scans and emails. Exams for remote students are proctored by the local faculty liaison. Exam papers are sent to the instructor by scans and emails too. All grading is either done by the instructor or the teaching assistant. Graded homework and exams are returned to the remote students by scans and emails.

\subsubsection{Case studies}

Towards the latter part of the semester, when students have been exposed and gained some experience with Bayesian inference, students are grouped to do case studies with real data applications. These case studies are all open-ended. Students are given the chance to freely explore the datasets and come up with their methods and realize their inferences through MCMC computation techniques. 

Prior to the case study class meeting, groups need to post their analyses onto the Moodle discussion forum to receive credit. Students in the same group take turn to be the leading writer of the analyses. Such practice ensures that everyone is prepared to discuss the approaches and findings from the group, and the class meetings usually turn out to be great discussion sessions and ideas bounce back and forth. 

\subsubsection{Final project}

As mentioned before, the course has a final project component, and students can choose from one of the following.

\begin{itemize}
\item[-] A Bayesian data analysis on a topic of your choosing.
\item[-] A new Bayesian methodology or theoretical finding.
\item[-] A Bayesian research paper or a book chapter (choose from a provided list).
\end{itemize}

Students submit project proposal after the first midterm exam. They are encouraged to meet with the instructor to discuss their project ideas and their progress along the way. 

The final project presentation has two parts: a 2-min video on Moodle, and a poster at the poster session. The choice of a poster session is to accommodate a relatively large class (16 local students). However, it mimics real research settings, as it has become common for academic conferences to have a poster session for graduate students and junior researchers. Students in the class overall enjoy being able to talk to audience as a small group. The amount of interaction between presenters and the audience is much more than a regular presentation.

\subsection{Five-Course MOOC}

\subsubsection{Quizzes}

Each module also features two sets of multiple choice quizzes, one formative and one summative. Each question is encoded with feedback that points learners back to relevant learning objectives. Learners can attempt the summative quizzes multiple times with slightly modified versions of the questions.

\subsubsection{Data analysis projects}

Each course ends with a data analysis project, the focus of which is summarized in Table \ref{tab:modules}, and the specialization wraps up with an extended capstone project. Each student who turns in a project evaluates three other students' work using a peer evaluation rubric. Learners are also strongly encouraged to seek informal feedback on their projects in the course discussion forums. All data analysis projects appearing in the courses in this specialization  are hosted in a publicly available GitHub repository at \href{github.com/StatsWithR/projects}{\url{https://github.com/StatsWithR/projects}}.

\section{Interaction}

\subsection{EDA Course}

Currently, there is limited interaction in this course.  Students communicate with the instructor by means of personal meetings or email or messages sent through the learning management system.  There is no regularly planned interaction between students such as an outline chat session, But students are asked to read and review the project presentations of two other students in the class.

\subsection{Bayesian Course}

To create and foster an online learning community, there are extensive uses of the online discussion forum on Moodle\footnote{Moodle is the course management system used at Vassar College.}. 

At the beginning of the semester, students make self-introduction posts about their basic information (name, year, and school), prior statistics exposure, prior R exposure and potential final project interests. 

During the semester, online discussion forums are created whenever sharing of information and making comments are needed, and they are for credit sometimes. For example, when covering the Gibbs sampler, the class reads a research paper Explaining the Gibbs Sampler by George Casella and Edward I. George \cite{CasellaGeorge1992}. A reading guide for this paper with 6 questions is provided to the students. Prior to the class meeting, students need to respond to any one question on the online discussion forum to receive participating credit. After the class discussion, students need to make another response for receive credit. Such requirement not only helps students in reading a statistics research paper outside of class, but also helps facilitate both in-class discussion by more engagement prior and post class meeting.

There is a final project for students in the course. In addition to present their projects in a poster session at the end of the semester, students need to make a 2-min video post about their projects on the online discussion forum. Watching a 2-min pitch talk prior to the poster session helps other students to arrange their poster visits. These videos also help students to succinctly present their projects in a manner as appealing as possible.

While the use of the online discussion forum on Moodle is motivated first mostly to make remote students feel included, it ultimately engages local students much better as well.

\subsection{Five Course MOOC}

Student interaction, or the lack thereof of in person interaction, is often a major challenge in online courses. However in MOOCs the discussion forums are often a major strength of the course. Given that at any point thousands of students are enrolled in the course, even if a small percentage of these students choose to browse the discussion forums, and an even smaller percentage of them interact with other learners on the course discussion forums, this still results in a large number of learners interacting with each other.

Additionally, over the years of the MOOC being offered, a handful of very knowledgeable and helpful course mentors emerged from the discussion forums. These are learners who took the courses at some point and now volunteer their time to answer student questions and provide direction for new learners.

\section{Challenges}

\subsection{EDA Course}

There are several current challenges in the current version of the EDA online course. 

\begin{itemize}

\item Interaction with the student

It is beneficial if the student can interact with the instructor and fellow students in an online course.  Some attempts for interaction such as online chat sessions or online message boards don't appear to be effective in this particular class.  So most of the communication is done through email and personal meetings.  There is an effort to try out new methods of interaction when they become available.

\item Technology issues

Students can get frustrated with technology issues such as installing software or getting their R markdown files to ``knit" properly.  It is best to address these issues early in the course so the course is more about the EDA concepts and less about the associated technology.  

\item Balance of computation and interpretation in assignment work

In a typical assignment, the student will turn in a Markdown file that blends output from the R system and written text that interprets the R output in the context of the particular applied problem.  Since the course is really focusing on the interpretation rather than the implementation of the EDA methods, the assignment should emphasize the interpretation component.  Depending on the background, the student may emphasize instead the computation component, but hopefully the students will learn what is expected in future assignments.

\end{itemize}

\subsection{Bayesian Course}

Although the instructor has been faced with various challenges, advice, suggestions, and sharing from multiple parties have been tremendously helpful. To improve the teaching and learning mode of the hybrid model, lecture videos can be edited and shortened when resources permit. If done properly, the lecture videos can well be a set of learning material for anyone (not necessarily from the LACOL member colleges) who is interested in an undergraduate-level introduction to Bayesian statistics. More thought and consideration on turning course material into video form can further enhance the teaching and learning. While students' interaction can be maintained using online discussion forum on Moodle, other forms of interaction can be explored and developed to further enhance the overall interaction in the course.

\subsection{Five Course MOOC}

There are three main challenges with offering this content on an online platform; two of these are associated with to the labs and the other is associated with the data analysis projects.

\begin{itemize}

\item Autograding: Given that this is a course with thousands of learners enrolled at any given point, human-grading is simply not feasible. The lab assessments are set up as multiple choice questions. Learners complete the lab exercises by generating R Markdown reports in which they analyze a dataset. Then, they answer a series of multiple choice questions about the data analysis results. The challenge is that the multiple choice questions do not assess the full spectrum of the skills we want learners' to acquire via these labs -- they assess whether they can obtain the correct results using R, however they do not assess mastery of R syntax, reproducibility of their analysis, etc. 

\item Computing infrastructure: Our preferred method for getting students with no computing background started with R is a cloud-based access to RStudio in order to avoid challenges around local installation and to provide a uniform computing environment for all learners. However it is not feasible to offer a centralized cloud-based solution to all learners enrolled in a MOOC, and hence students have to locally install R and RStudio and the correct versions of all packages they use in the labs.

As a partial solution for this challenge, we offer students the option to complete the labs in the first course of the specialization on DataCamp (\href{www.datacamp.com}{\url{https://www.datacamp.com}}), an online learning platform that provides in-browser access to RStudio. This helps students struggling with software installation issues early on in the course to get started with data analysis and go back to tackling software challenges once they feel a little bit more confident with R.

\item Peer evaluation: Autograding is not feasible for open ended data analysis projects, and hence peer evaluation is the only solution for grading of these projects. Even with a very detailed rubric, consistency in grading is difficult to attain, and it is challenging for learners who are just learning the material themselves to evaluate others' work. Additionally, variability in quality and depth of feedback provided leaves can leave learners frustrated. The option to share their projects on the discussion forums and get feedback can be helpful for some learners, but others are not so keen on publicly sharing their projects.

\end{itemize}

\section{Concluding Comments}

Although these is  current interest in teaching online introductory statistics courses, the online statistics courses described here are directed towards specific groups of students and all of the comments may not be directly applicable to the introductory class.   For example, the students in the EDA online course are primarily masters-level or advanced undergraduate students who are comfortable in working independently on assignments and projects, and this particular course design may not be suitable for an introductory statistics class with a minimal mathematics prerequisite.  But there are general common elements to these courses which would be helpful for the instructor who is designing the first online statistics course.

\subsection*{Presentation of Content}

Although there is some written instructional content in the EDA course, there is an extensive use of videos in all of these online courses.   The use of videos makes it possible for several instructors to get involved in the presentation of content.  In addition, it is possible for the student to learn from the video at his or her own pace, replaying parts of the video to help understand the material.

\subsection*{Interaction}

It is important to develop a cooperative learning environment among students in the online course.  These courses suggest some useful methods for facilitating this type of environment.  Discussion forums, as described by the Five-Course MOOC, are one good way to foster communication between students.  Another good opportunity for collaboration is through statistics projects where groups of students explore data on case studies.

\subsection*{Assessment}

It should be noted that traditional forms of assessment such as multiple choice exams play a limited role in assessment for these online statistics courses.  Instead, these courses feature data analysis projects, interactive computer lab assignments, and other projects where the student does an exploration into a new method or finding that is not covered in the curriculum.

\subsection*{Using Software}

All of these online courses use technology or software that may not familiar to the student.  Specifically, since these are statistics courses, one would typically use the R software together with some specialized R packages.  Introducing this technology creates some challenges since the students can vary greatly with their experience with the R system.  These courses have presented some ways to mitigate these technology challenges by creating special packages (such as the LearnEDA package in the EDA course) that include all of the datasets and special functions needed for the course.  The Five-Course MOOC provides some suggestions to help some of these technology issues such as using a cloud-based version of the R system or outsourcing some of the tutorial R material to a company such as DataCamp.  The instructor teaching an online statistics course should think carefully about the use of software, especially from the viewpoint of the student who is inexperienced with technology.

\bibliography{online-stats}{}
\bibliographystyle{ieeetr}

\end{document}